\documentclass[useAMS,usenatbib, a4paper]{mn2e}

\usepackage[dvips]{graphicx}
\usepackage{subfigure}
\usepackage{float}
\usepackage{amsmath}
\usepackage{color}

\newcommand{\apj}{ApJ}
\newcommand{\apjl}{ApJL}
\newcommand{\pasj}{PASJ}
\newcommand{\aap}{A\&A}

\newcommand{\aj}{AJ} 
\newcommand{\mnras}{MNRAS}

\title[Environmental dependence of star formation induced by cloud collisions in a barred galaxy]{Environmental dependence of star formation induced by cloud collisions in a barred galaxy}
\author[FUJIMOTO ET AL.]{Yusuke Fujimoto\thanks{E-mail:
yusuke@astro1.sci.hokudai.ac.jp}, Elizabeth J. Tasker and Asao Habe\\
Department of Physics, Faculty of Science, Hokkaido University, Kita 10 Nishi 8 Kita-ku, Sapporo 060-0810, Japan}
\begin{document}

\pagerange{\pageref{firstpage}--\pageref{lastpage}} \pubyear{2014}

\maketitle

\label{firstpage}

\begin{abstract}
Cloud collision have been proposed as a way to link the small-scale star formation process with the observed global relation between the surface star formation rate and gas surface density. We suggest that this model can be improved further by allowing the productivity of such collisions to depend on the relative velocity of the two clouds. Our adjustment implements a simple step function that results in the most successful collisions being at the observed velocities for triggered star formation. By applying this to a high resolution simulation of a barred galaxy, we successfully reproduce the observational result that the star formation efficiency (SFE) in the bar is lower than that in the spiral arms. This is not possible when we use an efficiency dependent on the internal turbulence properties of the clouds. Our results suggest that high velocity collisions driven by the gravitational pull of the clouds are responsible for the low bar SFE.

\end{abstract}


\section{Introduction}

Recent studies of galactic-scale star formation have revealed that the rate at which stars are produced depends on the galactic environment. Global structures within a galaxy result in changes to the star formation rates (SFR), even when the gas surface density is almost the same. In observations of disc galaxies, the star formation efficiency (${\rm SFE} = \Sigma_{\rm SFR} / \Sigma_{\rm gas}$) is found to vary between the nucleus and disc region \citep{Muraoka2007}, and between the arm and inter-arm \citep{Muraoka2009, Hirota2014}. 

High resolution ($\sim$ 250\,pc) $^{12}{\rm {CO}} (J = 1 - 0)$ observations of the barred galaxy M61 (NGC 4303) by \citet{Momose2010}, showed that the SFE in the bar is 50\,\% of that in the spiral arms. Previous studies have suggested that this drop is due to strong shear along the bar that provides a turbulence injection to support the giant molecular clouds (GMCs) \citep{Tubbs1982, Athanassoula1992, Downes1996, ReynaudDownes1998, Sorai2012, Meidt2013}. However, there is counter evidence suggesting that the role of shear is too small to be consequential to the evolution of the GMCs. For clouds in our own Galaxy, \citet{Dib2012} found that shear is consistently a fraction of the value needed to disrupt a density perturbation, and thereby does not affect star formation. 

A way to resolve this disparage is to use simulations, yet here too there is disagreement. In two-dimensional models of the barred galaxy M83, \citet{Nimori2012} found that the SFE in the bar region was 60 percent of the spiral arm due to the strong internal turbulence of the clouds. Conversely, more recent 3D models of the same galaxy performed by \citet{Fujimoto2014}, found that the typical internal cloud velocity dispersion showed little variation between clouds forming in the bar, spiral arm and disc. This implied shear might not be the key to understanding the varying star formation rate. Yet, the situation is complicated by the SFE being highly dependent on the stellar model used. 

\citet{Fujimoto2014} did not include active star formation, but estimated the SFE based on the gas properties. There are multiple methods for doing this, each based on assumptions as to what governs the production of stars. Two methods used by \citet{Fujimoto2014} compared a scheme utilising the internal properties of the cloud with one that considered star formation driven by cloud interactions. The former was proposed by \citet{KrumholzMcKee2005} and assumed that clouds are supersonically turbulent with a log-normal density distribution. By demanding that gas collapses when the gravitational energy within a cloud exceeds its turbulent energy, they find the SFR per cloud is:

\begin{equation}
{\rm SFR}_{\rm c} = \epsilon_{\rm ff}\ \left(\frac{\alpha_{\rm vir}}{1.3}\right)^{-0.68}\left(\frac{\cal{M}}{100}\right)^{-0.32}\frac{M_{\rm c}}{t_{\rm ff}}
\label{eq:stars_turbulence}
\end{equation}

\noindent where the star formation efficiency per free-fall time, $\epsilon_{\rm ff} = 0.014$, the virial parameter $\alpha_{\rm vir} = 5\sigma_{\rm c}^2R_{\rm c}/GM_{\rm c}$, properties $R_{\rm c}$, $M_{\rm c}$ and $\sigma_{\rm c} $ are the cloud radius, mass and 1D velocity dispersion, the Mach number is the ratio between the cloud's velocity dispersion and the sound speed, $\cal{M}$ $\equiv \sigma_{\rm c}/c_{\rm s}$, and $t_{\rm ff}$ is the cloud free-fall time.

The second scheme assumes that star formation is initiated by collisions between clouds. Such interactions can trigger a shock at the collision interface which fragments into stars. This has been suggested as a way to unite the local star formation process with the globally observed Kennicutt-Schmidt relation \citep{Tan2000, TaskerTan2009, Kennicutt1998} and also as a mechanism to create massive stars and super star clusters \citep{Takahira2014, Fukui2014,  Dib2013, Furukawa2009, Ohama2010, HabeOhta1992}. Since there is evidence that clouds may be gravitationally unbound \citep{Heyer2009, Dobbs2011}, cloud collisions could be the best candidate to create the dense regions where stars are formed. In the shock-generating model proposed by \citet{Tan2000}, the surface density of the SFR becomes:

\begin{equation}
\Sigma_{\rm SFR} = \frac{\epsilon f_{\rm sf} N_{\rm A} M_{\rm c}}{t_{\rm coll}}
\label{eq:cloud-cloud collision}
\end{equation}

\noindent where $\epsilon = 0.2$ is the total fraction of cloud gas converted to stars during a star-forming collision, $f_{\rm sf}$ is the fraction of cloud collisions which successfully lead to star formation, $N_{\rm A}$ is the surface number density of clouds, $M_{\rm c}$ is the cloud mass and $t_{\rm coll}$, the typical time between collisions. 

\citet{Fujimoto2014} found that neither of these models produced the observed lower SFE in the bar region. With the turbulence model in Eq.~\ref{eq:stars_turbulence}, the SFE was equal in the bar and spiral arms, while the cloud collision model in Eq.~\ref{eq:cloud-cloud collision} gave a higher SFE in the bar than the arms. The first of these was consistent with the result that the typical cloud properties were not dependent on cloud environment. The cloud collision model reflected the fact that the collision rate was highest in the densely packed bar environment, leading to a higher star formation rate. In this case, the difference in the galactic environment produced the opposite effect to that in observations, but this might have been due to a simplification made in Eq.~\ref{eq:cloud-cloud collision}.

The cloud collision model assumes that all collisions are equally likely to produce star formation. The fractional success rate, $f_{\rm sf}$, is taken to be constant whose value was selected to be 0.5 in \citet{Fujimoto2014}. This approximation is known to be inaccurate. In observations of triggered star formation in cloud collisions by \citet{Fukui2014, Ohama2010, Furukawa2009}, the formation of super star clusters was found to be associated with collisional velocities around 20\,km/s. This was investigated in simulations by \cite{Takahira2014}, who found that the production of star-forming cores in a collision strongly depended on the relative velocity of the two clouds. A slow collision would not produce a shock strong enough to create a dense region at the cloud interface while too fast a collision would result in the shock front exiting the cloud before a core had time to form. The value of $f_{\rm sf}$ should therefore depend on the relative velocity of the colliding clouds. 

To be strictly accurate, the values for $f_{\rm sf}$ and $\epsilon$ are not independent. If $f_{\rm sf}$ depends on velocity, then a prime velocity for production of stars is going to produce a higher efficiency. However, for the sake of simplicity, we consider $\epsilon$ to represent the average conversion efficiency and assume only $f_{\rm sf}$ varies. We keep the value of $\epsilon$ suggested in \citet{Tan2000} as $0.2$.

In this letter, we calculate the SFE in the barred galaxy model of \citet{Fujimoto2014} using a revised version of the cloud collision star formation scheme in Eq.~\ref{eq:cloud-cloud collision}, accounting for variations in stellar production due to the collisional velocity of the clouds. We compare the results in different galactic regions of the simulation with those in observations.

\section{Simulation}
The simulation presented in this letter was run using {\it Enzo}: a 3D adaptive mesh refinement (AMR) hydrodynamics code \citep{Enzo}. Our galaxy was modelled on the nearby barred spiral galaxy, M83, with the gas distribution and stellar potential taken from observational results. For the galaxy's radial gas distribution, we assumed an initial exponential density profile based on the observations of \citet{Lundgren2004} and used fixed-motion star particles to create a stellar potential with the observed pattern speed for M83.

The GMCs in our simulation were identified as coherent structures contained within contours at a threshold density of $n_{\rm H,c} = 100$\,cm$^{-3}$, similar to the observed mean volume densities of typical galactic giant molecular clouds. The simulation outputs were analysed every 1\,Myr and the clouds mapped between consecutive times. This cloud tracking algorithm is described in \citet{TaskerTan2009}. A collision is said to have happened when a single cloud is at the predicted position for two other clouds after 1\,Myr of evolution. It is a lower estimate of the true cloud interaction rate, since it does not include tidal shredding where two identifiable objects exist at the end of the encounter.

After 120\,Myr (one pattern rotation period), the gas is fully fragmented, and between 200\,Myr and 280\,Myr, the galactic disc settles into a quasi-equilibrium with no large structural change. We analysed the clouds at 240 Myr.

We assigned an environment group based on the cloud's physical location within the disc. If a cloud is found within the galactic radii $2.5 < r < 7.0$\,kpc, it is recognised as a spiral cloud. Outside $r = 7.0$\,kpc, clouds are designated disc clouds. Bar clouds form in a box-like region at the galactic centre, with a length of 5.0\,kpc and width 1.2\,kpc. The nucleus region inside 600\,pc is excluded due to the difficulty in accurately tracking clouds in such a high density area. The boundaries of these three regions are shown in Figure 5 of \citet{Fujimoto2014}, along with a more thorough description of the simulation.

\section{Results}

\subsection{Environmental dependence of collision velocity}

\begin{figure}
\begin{center}
	\includegraphics[width=80mm]{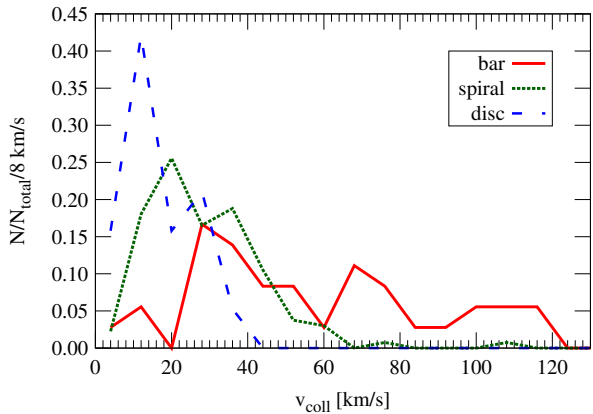}
	\caption{Distribution of the collision velocity of clouds colliding in the bar region (solid red line), spiral region (dotted green line) and disc region (dashed blue line). The collision events are counted for 10\,Myrs from $t =$ 230 Myr to 240 Myr.}
	\label{distribution of collision velocity classified by galactic regions}
\end{center}
\end{figure}

\begin{figure}
\begin{center}
	\includegraphics[width=80mm]{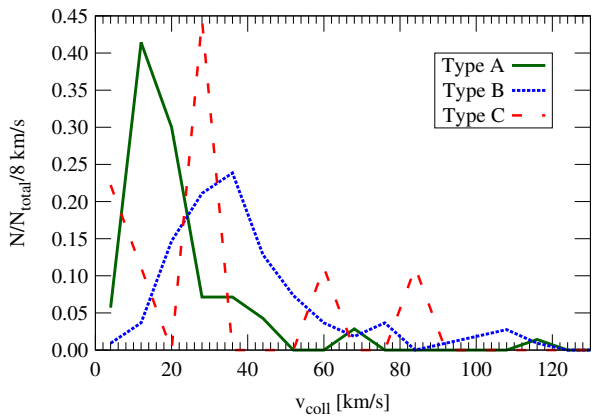}
	\caption{Distribution of the collision velocity for three cloud categorisations: {\it Type A} (solid green line) are the most common clouds with properties typical of observed GMCs, {\it Type B} (dotted blue line) are massive associations and {\it Type C} (dashed red line) are unbound, transient clouds. The cloud type plotted is that of the most massive cloud involved in the collision.}
	\label{distribution of collision velocity classified by cloud type}
\end{center}
\end{figure}

In order to determine whether collision velocity is likely to play a role in the productiveness of triggered star formation, we looked at the range of values present in the simulation. Figure~\ref{distribution of collision velocity classified by galactic regions} shows the collision velocity distribution in each of our three galactic region. This was calculated from the relative velocity of the clouds 1\,Myr prior to their merger, $v_{\rm coll} = |\overrightarrow{v_1} - \overrightarrow{v_2}|$, where  $\overrightarrow{v_1}$ and $\overrightarrow{v_2}$ are the bulk velocities of the clouds. Since the number of collisions occurring in a 1\,Myr output interval is small, we included all collisions occurring between $t = 230 - 240$\,Myr, during the quiescent period of the disc's evolution.

The collision velocity shows a clear dependence on the galactic environment. The bar region has the widest range, extending out to 120\,km/s with only a small peak in the distribution around 35\,km/s. By contrast, the spiral region has a steeped profile with a typical collision speed of 20\,km/s and clouds rarely colliding faster than 60\,km/s. In the disc region, interactions are more gentle with a collision speed peaking at only 15\,km/s and a maximum of 40\,km/s.

By virtue of its area, the total number of cloud collisions is highest in the spiral region. However, the fraction of clouds undergoing a collision event is higher in the bar, due to the constrained elliptical motion. This bolstered rate of interaction alters the properties of the clouds, allowing much larger structures to be built through successive mergers. The result is a group of Giant Molecular Associations (GMAs) that dominate the local gravitational field, pulling clouds onto a collision course that increases their relative velocity.

That the collision velocity can be attributed to properties of the cloud population can be seen clearly in Figure~\ref{distribution of collision velocity classified by cloud type}. The collision velocity is now plotted for three different types of cloud, categorised by their mass and radius. {\it Type A} clouds form the main population in the disc. Their properties are consistent with observed GMCs with masses $\sim 5\times 10^5$\,M$_\odot$ and radii $\sim 11$\,pc. {\it Type B} clouds are the giant GMAs with masses above $10^6$\,M$_\odot$. {\it Type C} clouds are unbound objects that have short lives in filaments and the tidal tails of larger interacting clouds. The cloud type was determined using the mass-radius relation described in detail in \citet{Fujimoto2014} and the collision cloud type is dictated by the largest object of the interacting pair, resulting in a low number of {\it Type C} with a noisy distribution. These distributions show that collisions involving a GMA {\it Type B} cloud are faster. The speed peaks around 40\,km/s and extends to beyond 100\,km/s. By contrast, a typical collisions with a {\it Type A} cloud occurs at half the speed of those with {\it Type B}. This fast interaction speed for the {\it Type B} is consistent with their escape velocity, which ranges between $17 - 100$\,km/s for the clouds with masses between $10^6 - 10^8$\,M$_\odot$ and radii $30 - 100$\,pc. The {\it Type B} clouds form a dominant population in the bar as shown in Table~\ref{table:cloud_percentage}. Half the clouds in the bar are either {\it Type B} or the small {\it Type C} forming in the interaction tails of the {\it Type B}. The spiral and disc region have a stronger population of {\it Type A} clouds, with the disc having less than 6\% {\it Type B}. The {\it Type B} clouds therefore govern the interactions in the bar, producing a higher collision speed than either the spiral or the disc. This supports the idea that the variation in SFE with galactic region might be due to a variation in the efficiency of triggered star formation.

\begin{table}
\begin{center}
\begin{tabular}{|l|c|c|c|} \hline
	& bar & spiral & disc \\ \hline
	Type A & 49.4\% (38/77) & 64.1\% (330/515) & 83.3\% (85/102) \\ \hline
	Type B & 13.0\% (10/77) & 12.8\% (66/515) & 5.9\% (6/102) \\ \hline
	Type C & 37.7\% (29/77) & 23.1\% (119/515) & 10.8\% (11/102) \\ \hline
\end{tabular} 
\caption{The percentage of each cloud type in each galactic region at $t = 240$\,Myr. Bracketed numbers show the actual number of clouds of that type divided by the total cloud number in the region.}
\label{table:cloud_percentage}
\end{center}
\end{table}

\subsection{Environmental dependence of SFE}

\begin{table}
\begin{center}
\begin{tabular}{|l|c|c|c|} \hline
	$v_{\rm coll}$ & 0 $\sim$ 10 km/s & 10 $\sim$ 40 km/s & 40 km/s $\sim$ \\ \hline
	$f_{\rm sf}$ & 0.05 & 0.5 & 0.05 \\ \hline
\end{tabular} 
\caption{The fraction of collisions that result in star formation, $f_{\rm sf}$, for different collision velocities, $v_{\rm coll}.$}
\label{f_sf}
\end{center}
\end{table}

\begin{figure*}
\begin{center}
\subfigure{
	\includegraphics[width=85mm]{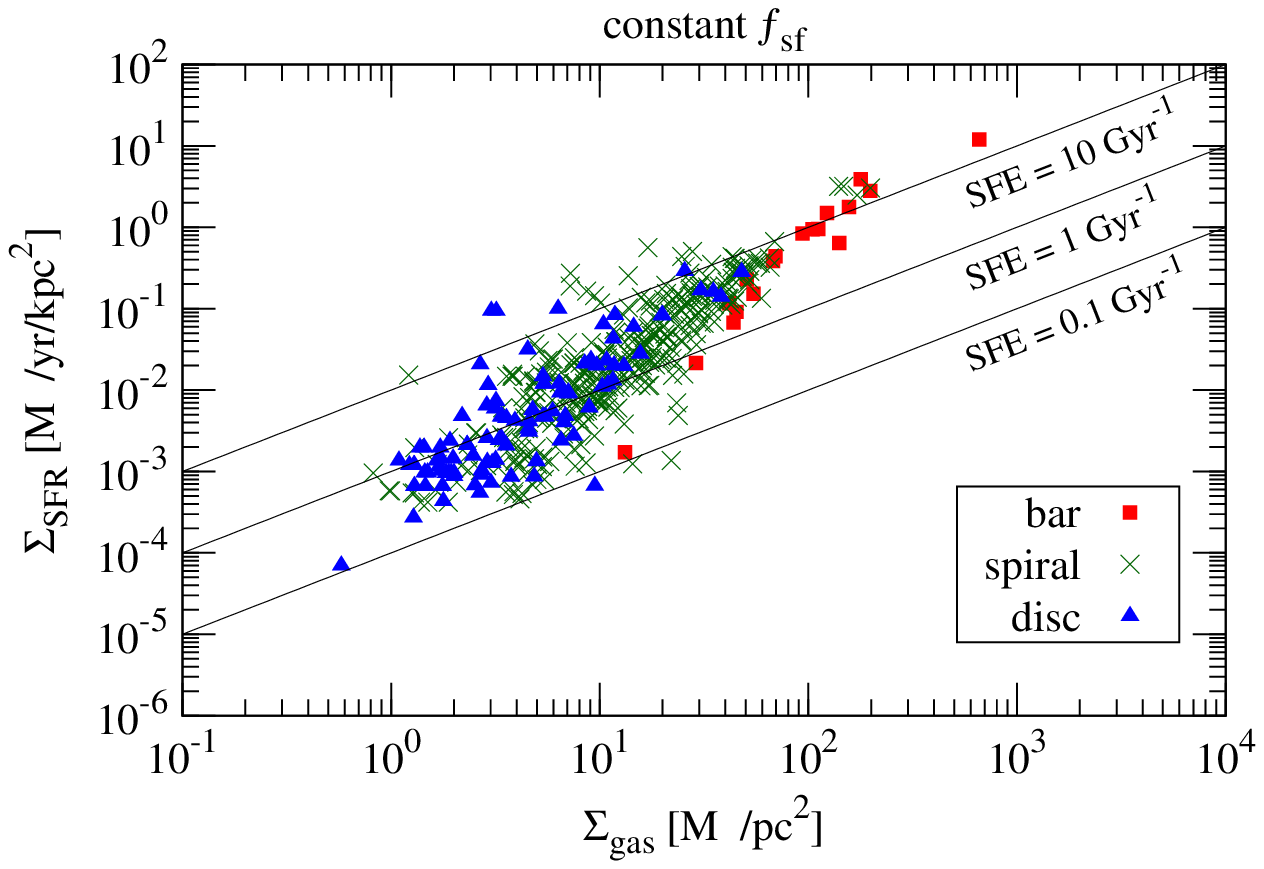}
	\includegraphics[width=85mm]{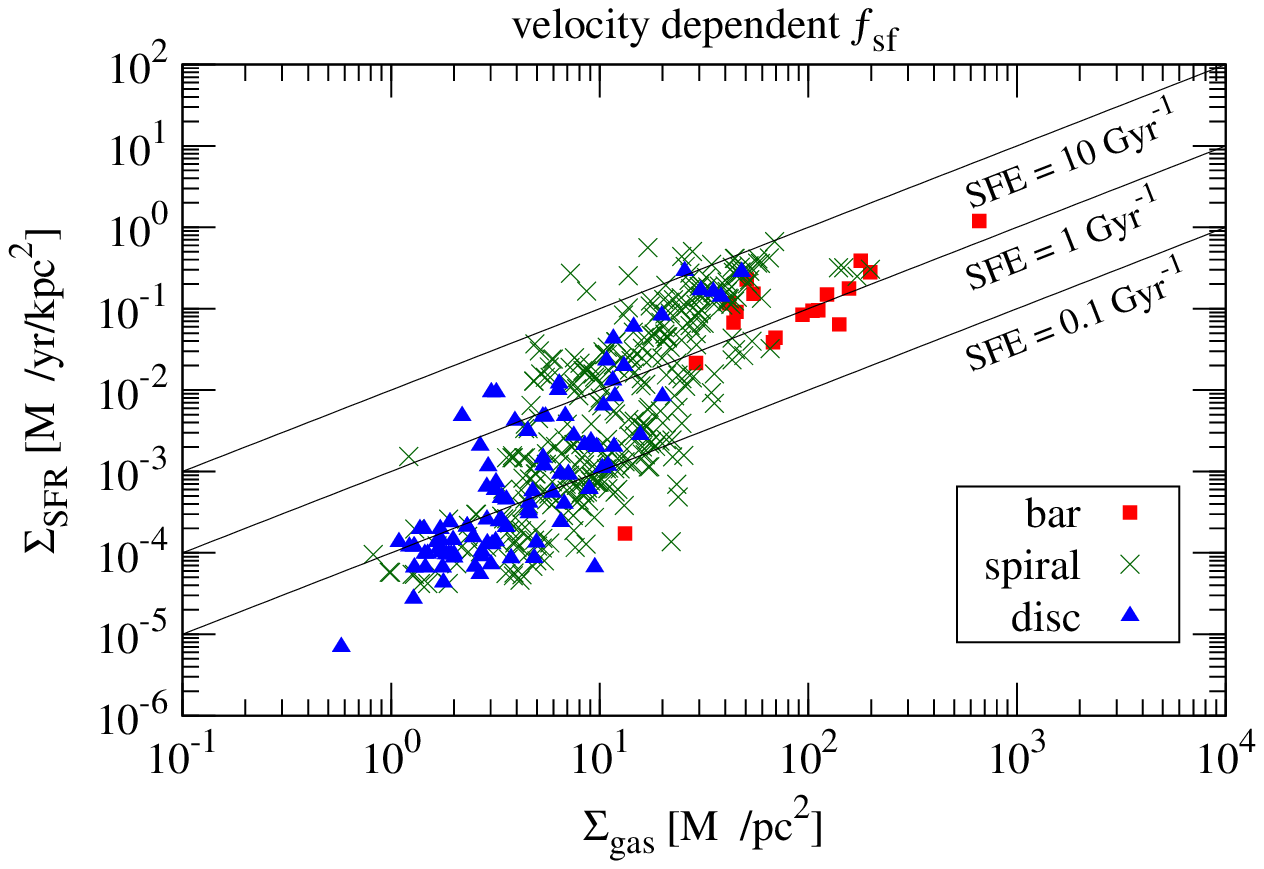}}
	\caption{Relation between gas surface density and SFR surface density. The SFR is estimated by the cloud collision model depending on the average collision velocity in 500\,pc $\times$ 5\,kpc cylindrical regions. The left panel uses a constant $f_{\rm sf}$ while the right panel uses the collision velocity dependent $f_{\rm sf}$, as described in Table~\ref{f_sf}. The coloured markers denotes different galactic regions: red squares are clouds in the bar region, green $\times$ are for the spiral region and blue triangles are those in the disc region. The black dotted lines show constant SFEs = $10,\ 1,\ 0.1\ \rm Gyr^{-1}$.}
	\label{surface density vs SFR}
\end{center}
\end{figure*}

To equate the cloud collision rate to a star formation rate, we used the model proposed by \citet{Tan2000} in Eq.~\ref{eq:cloud-cloud collision} to plot the observed Kennicutt-Schmidt relation between the surface star formation rate density and surface gas density. We calculated the surface area in the $x-y$ plane (face-on disc) and averaged the data over cylindrical regions with height 5\,kpc and radii 500\,pc, in keeping with the typical observational resolution. If we assume a constant fraction of all collisions lead to star formation with $f_{\rm sf} = 0.5$ in Eq.~\ref{eq:cloud-cloud collision}, we get the Kennicutt-Schmidt relation in the left-hand panel of Figure~\ref{surface density vs SFR}. The SFE is proportional to the frequency of collisions, with the bar having the highest number of cloud collisions and therefore the highest efficiency. 

In order to take into account the dependence of $f_{\rm sf}$ on the collision velocity, we use the step function described in Table~\ref{f_sf}. Since observations of triggered star formation found a collision velocity around $\sim 20$\,km/s \citep{Fukui2014, Furukawa2009, Ohama2010}, we assume that collisions between $10-40$\,km/s are ideally suited to creating stars and set $f_{\rm sf} = 0.5$ in this range. This is consistent with the majority of our collisions in Figure~\ref{distribution of collision velocity classified by galactic regions}; a necessity since the average SFR is in good agreement with observations \citep{Bigiel2008}. Outside this region, we assume that a collision velocity less than 10\,km/s is too slow to form the compressed shock front that leads to efficient star formation while velocities higher than 40\,km/s produce interaction times between clouds that are too short to form a high number of stars. Within these ranges, we therefore lower the value of $f_{\rm sf}$ by a factor of 10. Exactly  where these cut-off should be and the correct value for $f_{\rm sf}$ requires a further exploration of shock speed and cloud structure, so here we select values that lie within our observed range of velocities and agreed with the observations available. 

The result of including this collision velocity dependence is shown in the right-hand panel of Figure~\ref{surface density vs SFR}. With a significant fraction of the bar clouds colliding at high velocities with a low success rate, the SFE in this region drops. This produces a lower SFE in the bar than a large fraction of the spiral regions, in agreement with the observations of \citet{Momose2010}, where the bar SFE lies in the lower-half of the SFE spread in the spiral. The spiral region shows an increased range of efficiencies due to its broad profile of collision velocities in Figure~\ref{distribution of collision velocity classified by galactic regions} having tails in both our low efficiency regions. Conversely most of the disc clouds collide too slowly for strong star formation, decreasing the efficiency of the majority of these points. This demonstrates that while the cloud collision model for star formation can successfully reproduce the Kennicutt-Schmidt relation, the dependence on the velocity of the interaction introduces an environmental dependence that cannot be ignored.

\begin{figure}
\begin{center}
	\includegraphics[width=80mm]{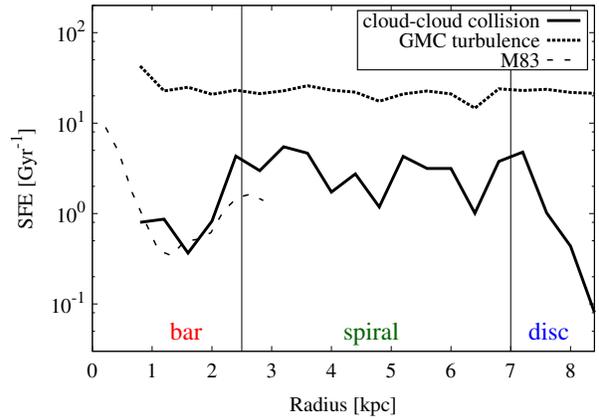}
	\caption{Azimuthally averaged (bin size 400 pc) radial distributions of the SFE estimated by the velocity dependent cloud collision model (solid line), the cloud turbulence model (dashed line) and from observations of M83 by \citet{Hirota2014} (wide dashes). The vertical lines show the borders of galactic regions: $600 {\rm pc} < r < 2.5\ \rm kpc$ marks the bar region, $2.5 < r < 7.0\ \rm kpc$ is the spiral region, and $r > 7.0\ \rm kpc$ is the disc region.}
	\label{radii vs SFE}
\end{center}
\end{figure}

A second comparison point is the radial distribution of the SFE was measured by \citet{Hirota2014} for the bar region in M83. They found a peak in the SFE at the bar end, corresponding to an efficiency of $\sim 2$\,Gyr$^{-1}$ which drops to $\sim 0.3$\,Gyr$^{-1}$ along the bar length. We compared this to the radial profiles we get using the cloud collision and turbulence models for SFE in Figure~\ref{radii vs SFE}. For star formation estimated from the cloud turbulence in Eq.~\ref{eq:stars_turbulence}, there is no radial dependence and we see a constant efficiency through the three galaxy environments. Using the velocity dependent collision scheme, a clear drop is seen after the bar end of the same magnitude as in the observations.

\section{Conclusion}

We investigated the variation in the relative velocity of cloud collisions in different regions of a simulated galaxy. Using this, we proposed a change to the triggered star formation model developed by \citet{Tan2000} that varied the effectiveness of star formation from cloud collisions based on the speed at which the clouds collide. Taking the observations of triggered star formation as the most successful velocity for forming stars, we varied the fraction of collisions that would result in star formation such that collisions between $10-40$\,km/s were successful 50\,\% of the time, while collisions slower and faster than this range were only successful 5\,\% of the time. Our main results are:

\begin{itemize}
\item The collision velocity shows a clear dependence on galactic environment. Clouds formed in the bar region typically collide faster than those in the spiral. Clouds in the disc are involved in the gentlest collisions.

\item This dependence is due to the distribution of cloud types in the three regions. A higher fraction of bar clouds are massive, creating strong gravitational interactions that increase the collision velocity. 

\item The unproductive collisions in the bar region lowers the SFE to put it below the maximum efficiency in the spiral region. This is despite collisions being more common in the bar region. The result is in agreement with observations of other barred galaxies.

\item When plotted as a radial distribution, the drop in SFE at the bar end is also consistent with observational measurements of M83. This is not reproduced when the SFE is calculated from the cloud turbulence, which shows a constant SFE across all regions.

\end{itemize}

One area not considered is the effect of stellar feedback. The impact of feedback is a hot topic, with opinions suggesting that it does not affect the GMCs \citep{Renaud2013} to implying it is a controlling force. In stellar wind models performed by \citet{Dib2013}, the SFE in a star-forming clump decreased with clump mass. Such an effect could also explain the lower bar SFE if the larger GMAs resulted in more massive clumps. However, the authors also find that for star clusters thought to be formed in cloud collisions, the best-fit model uses an increasing SFR with time. This is likely to be controlled by the evolution of the collision-induced surface density, which \citet{Takahira2014} demonstrated depends strongly on collision speed. The two effects will therefore work in parallel and further investigation is needed to separate out their contributions.

\section*{Acknowledgments}

The authors would like to thank the yt development team \citep{Turk2011} for help during the analysis of these simulations. Numerical computations were carried out at the Center for Computational Astrophysics (CfCA) of the National Astronomical Observatory of Japan. EJT is funded by the MEXT grant for the Tenure Track System and the Suhara Memorial Foundation.

\bsp

\label{lastpage}


\begin{thebibliography}{99}
\bibitem[\protect\citeauthoryear{Athanassoula}{1992}]{Athanassoula1992} Athanassoula, E.\ 1992, \mnras, 259, 345 
\bibitem[Bigiel et al.(2008)]{Bigiel2008} Bigiel, F., Leroy, A., Walter, F., Brinks, E., de Blok, W.~J.~G., Madore, B., \& Thornley, M.~D.\ 2008, \aj, 136, 2846 
\bibitem[\protect\citeauthoryear{Daddi et al.}{2010}]{Daddi2010} Daddi, E., Elbaz, D., Walter, F., et al.\ 2010, ApJ, 714, L118
\bibitem[Dib et al.(2012)]{Dib2012} Dib, S., Helou, G., Moore, T.~J.~T., Urquhart, J.~S., \& Dariush, A.\ 2012, \apj, 758, 125 
\bibitem[Dib et al.(2013)]{Dib2013} Dib, S., Gutkin, J., Brandner, W., \& Basu, S.\ 2013, \mnras, 436, 3727 
\bibitem[\protect\citeauthoryear{Dobbs et al.}{2011}]{Dobbs2011} Dobbs, C.~L., Burkert, 
A., \& Pringle, J.~E.\ 2011, \mnras, 413, 2935 
\bibitem[\protect\citeauthoryear{Downes et al.}{1996}]{Downes1996} Downes, D., Reynaud, D., Solomon, P.~M., \& Radford, S.~J.~E.\ 1996, \apj, 461, 186 
\bibitem[\protect\citeauthoryear{Fujimoto et al.}{2014}]{Fujimoto2014} Fujimoto, Y., Tasker, E.~J., Wakayama, M., \& Habe, A.\ 2014, \mnras, 439, 936
\bibitem[\protect\citeauthoryear{Fukui et al.}{2014}]{Fukui2014}  Fukui, Y., Ohama, A., Hanaoka, N., et al.\ 2014, \apj, 780, 36 
\bibitem[Furukawa et al.(2009)]{Furukawa2009} Furukawa, N., Dawson, J.~R., Ohama, A., et al.\ 2009, \apjl, 696, L115
\bibitem[\protect\citeauthoryear{Habe \& Ohta}{1992}]{HabeOhta1992} Habe, A., \& Ohta, K.\ 1992, PASJ, 44, 203 
\bibitem[\protect\citeauthoryear{Heyer et al.}{2009}]{Heyer2009} Heyer, M., Krawczyk, C., Duval, J., \& Jackson, J.~M.\ 2009, ApJ, 699, 1092 
\bibitem[\protect\citeauthoryear{Hirota et al.}{2014}]{Hirota2014} Hirota, A., Kuno, N., Baba, J., et al.\ 2014, \pasj, 66, 46
\bibitem[Kennicutt(1998)]{Kennicutt1998} Kennicutt, R.~C., Jr.\ 1998, \apj, 498, 541 
\bibitem[\protect\citeauthoryear{Krumholz \& McKee}{2005}]{KrumholzMcKee2005} Krumholz, M.~R., \& McKee, C.~F.\ 2005, ApJ, 630, 250
\bibitem[\protect\citeauthoryear{Leroy et al.}{2013}]{Leroy2013} Leroy, A.~K., Walter, F., 
Sandstrom, K., et al.\ 2013, \aj, 146, 19 
\bibitem[\protect\citeauthoryear{Lundgren et al.}{2004}]{Lundgren2004}Lundgren, A.~A., Wiklind, T., Olofsson, H., \& Rydbeck, G.\ 2004, A\&Ap, 413, 505
\bibitem[Meidt et al.(2013)]{Meidt2013} Meidt, S.~E., Schinnerer, E., Garc{\'{\i}}a-Burillo, S., et al.\ 2013, \apj, 779, 45
\bibitem[\protect\citeauthoryear{Momose et al.}{2010}]{Momose2010} Momose, R., Okumura, S.~K., Koda, J., \& Sawada, T.\ 2010, ApJ, 721, 383
\bibitem[\protect\citeauthoryear{Muraoka et al.}{2007}]{Muraoka2007} Muraoka, K., Kohno, K., 
Tosaki, T., et al.\ 2007, \pasj, 59, 43 
\bibitem[\protect\citeauthoryear{Muraoka et al.}{2009}]{Muraoka2009}Muraoka, K., Kohno, K., 
Tosaki, T., et al.\ 2009, \apj, 706, 1213  
\bibitem[\protect\citeauthoryear{Nimori et al.}{2012}]{Nimori2012} Nimori, M., Habe, A., Sorai, K., et al.\ 2012, MNRAS, 393 
\bibitem[Ohama et al.(2010)]{Ohama2010} Ohama, A., Dawson, J.~R., Furukawa, N., et al.\ 2010, \apj, 709, 975
\bibitem[Renaud et al.(2013)]{Renaud2013} Renaud, F., Bournaud, F., Emsellem, E., et al.\ 2013, \mnras, 436, 1836 
\bibitem[\protect\citeauthoryear{Reynaud \& Downes}{1998}]{ReynaudDownes1998} Reynaud, D., \& Downes, D.\ 1998, \aap, 337, 671 
\bibitem[\protect\citeauthoryear{Sorai et al.}{2012}]{Sorai2012} Sorai, K., Kuno, N., Nishiyama, K., et al.\ 2012, \pasj, 64, 51
\bibitem[\protect\citeauthoryear{Takahira et al.}{2014}]{Takahira2014} Takahira, K., Tasker, E.~J., \& Habe, A.\ 2014, arXiv:1407.4544 
\bibitem[\protect\citeauthoryear{Tan}{2000}]{Tan2000} Tan, J.~C.\ 2000, ApJ, 536, 173 
\bibitem[\protect\citeauthoryear{Tasker \& Tan}{2009}]{TaskerTan2009} Tasker, E.~J., \& Tan, J.~C.\ 2009, ApJ, 700, 358
\bibitem[\protect\citeauthoryear{The Enzo Collaboration et al.}{2013}]{Enzo} The Enzo Collaboration, Bryan, G.~L., Norman, M.~L., et al.\ 2013, arXiv:1307.2265
\bibitem[\protect\citeauthoryear{Tubbs}{1982}]{Tubbs1982} Tubbs, A.~D.\ 1982, \apj, 255, 
458 
\bibitem[Turk et al.(2011)]{Turk2011} Turk, M.~J., Smith, B.~D., Oishi, J.~S., et al.\ 2011, ApJS, 192, 9 

\end{thebibliography}
\end{document}